
\input harvmac
%
%
%

\def\tilde{\widetilde}
\def\bar{\overline}
\def\hat{\widehat}
\def\*{\star}
\def\[{\left[}
\def\]{\right]}
\def\({\left(}		
\def\){\right)}

%
%
\def\zb{{\bar{z} }}
\def\frac#1#2{{#1 \over #2}}
\def\inv#1{{1 \over #1}}

\def\d{\partial}

\def\2pi{\hbox{$2\pi i$}}

\def\dsl{\raise.15ex\hbox{/}\kern-.57em\partial}
\def\Dsl{\,\raise.15ex\hbox{/}\mkern-.13.5mu D}
%
%
\def\th{\theta}

\def\al{\alpha}

\def\sig{\sigma}	

%
%

%

\def\2pi{\hbox{$2\pi i$}}

\def\dsl{\raise.15ex\hbox{/}\kern-.57em\partial}
\def\Dsl{\,\raise.15ex\hbox{/}\mkern-.13.5mu D}
%
%
%
\font\numbers=cmss12
\font\upright=cmu10 scaled\magstep1
\def\stroke{\vrule height8pt width0.4pt depth-0.1pt}
\def\topfleck{\vrule height8pt width0.5pt depth-5.9pt}
\def\botfleck{\vrule height2pt width0.5pt depth0.1pt}
\def\Zmath{\vcenter{\hbox{\numbers\rlap{\rlap{Z}\kern
0.8pt\topfleck}\kern
2.2pt
                   \rlap Z\kern 6pt\botfleck\kern 1pt}}}
\def\Qmath{\vcenter{\hbox{\upright\rlap{\rlap{Q}\kern
                   3.8pt\stroke}\phantom{Q}}}}
\def\Nmath{\vcenter{\hbox{\upright\rlap{I}\kern 1.7pt N}}}
\def\Cmath{\vcenter{\hbox{\upright\rlap{\rlap{C}\kern
                   3.8pt\stroke}\phantom{C}}}}
\def\Rmath{\vcenter{\hbox{\upright\rlap{I}\kern 1.7pt R}}}
\def\Z{\ifmmode\Zmath\else$\Zmath$\fi}
\def\Q{\ifmmode\Qmath\else$\Qmath$\fi}
\def\N{\ifmmode\Nmath\else$\Nmath$\fi}
\def\C{\ifmmode\Cmath\else$\Cmath$\fi}
\def\R{\ifmmode\Rmath\else$\Rmath$\fi}

\Title{CLNS 95/1321, ~~ hep-th/9503088}
{\vbox{\centerline{Boundary Sine-Gordon Interactions at the   }
\centerline{ Free Fermion Point} }}

\bigskip
\bigskip

\centerline{M. Ameduri, R. Konik and A. LeClair}
\medskip\centerline{Newman Laboratory}
\centerline{Cornell University}
\centerline{Ithaca, NY  14853}
\bigskip\bigskip

\vskip .3in

We study bosonization of the sine-Gordon theory in the presence of
boundary interactions at the free fermion point.
In this way we obtain the boundary S-matrix as a function of
physical parameters in the boundary sine-Gordon Lagrangian.
The boundary S-matrix can be matched onto the solution of
Ghoshal and Zamolodchikov, thereby relating the formal
parameters in the latter solution to the physical parameters
in the lagrangian.

\Date{3/95}
%
%
%
%
%
%
\noblackbox

\def\phib{\bar{\phi}}
\def\zb{{\bar{z}}}

%
%
%
%
%
%
%
%
%
%

\def\bh{\hat{\beta}}

\newsec{Introduction}

It is now understood that for the integrable quantum field
theories in two dimensions, certain boundary interactions which
preserve the integrability are possible\ref\rgz{S. Ghoshal and
A. Zamolodchikov, Int. J. Mod. Phys. A9 (1994) 3841.}.
Apart from the theoretical interest, these models have interesting
applications  for example to
1D impurity problems \ref\rkf{C. Kane and M. Fisher,
Phys. Rev. B46 (1992) 15233}\ and edge excitations in fractional
quantum Hall states \ref\rwen{X. G. Wen,
Phys. Rev. B41 (1990) 12838.}\ref\rfls{P. Fendley,
A. W. W. Ludwig and H. Saleur,  cond-mat/9408068}.

In \rgz, boundary scattering matrices were computed for the sine-Gordon
model.  There, a general solution to the boundary Yang-Baxter equation
and the crossing-unitarity conditions was obtained which depends on two
formal parameters.  An important open problem in this work was the
relation of these formal parameters to the physical parameters in
the lagrangian.  In this paper we obtain this relation at the
free fermion point of the sine-Gordon theory, and see that
even here this relation is somewhat non-trivial\foot{After this work
was completed we learned that similar results were obtain in
\ref\rluca{L. Mezincescu, R. Nepomechie and A. Zamolodchikov,
in preparation.}.}.  (In applications to the fractional quantum Hall
effect, the free fermion point corresponds to filling fraction $\nu = 1/2$.)
Our computation involves understanding bosonization in the presence of
boundary interactions, which to our knowledge has not been
studied before.

\def\phib{\bar{\phi}}

\bigskip

\newsec{Bosonization with Boundary Interactions}

The sine-Gordon theory with an integrable boundary interaction is defined
by the Euclidean action
\eqn\eIIi{
S = \inv{4\pi}  \int dx dt
\inv{2} \( \d_z \Phi \d_\zb \Phi - 4 \lambda \cos \bh \Phi \)
- \frac{g}{4\pi} \int dt \cos \( \bh  (\Phi - \phi_0 ) / 2 \) , }
where
$z = (t+ix)/2 $, $\zb = (t-ix)/2 $.  Here, $\lambda$ and $g$ are
dimensionful parameters, $\bh$ is the coupling
and $\phi_0$ is a constant parameter\foot{
$\bh$ is related
to the conventional coupling $\beta$ by $\bh = \beta / \sqrt{4\pi}$,
and $\phi_0$ in this paper differs from the conventions in
\rgz\ by $\phi_0 \to \sqrt{4\pi} \phi_0$.
}.
The boundary is taken to be the time axis $-\infty < t < \infty$
at $x=0$.

One may view the above action as a boundary perturbation of a conformal
field theory.  Conformal field theories with boundary conditions that
preserve conformal symmetry were studied by Cardy\ref\rcardy{J. Cardy,
Nucl. Phys. B240 (1984) 514; Nucl. Phys. B324 (1989) 581.}.
Setting $g=\lambda = 0$, and requiring the boundary terms to vanish
in the variation of the resulting conformal action yields the
`free' boundary condition: $\d_z \Phi = \d_\zb \Phi $  at $x=0$.
In the conformal limit, $\Phi = \phi (z) + \phib (\zb ) $, and this
implies
\eqn\eIIii{
{\rm free}: ~~~~~~ \phi (z) =  \phib (\zb )  - \sigma  ~~~~~~(x=0),}
where $\sigma$ is a constant which represents data not
specified by the action
\eIIi.

Consider now the limit $g\to \infty$, where the bulk is still conformal
with $\lambda = 0$.  In this limit, vanishing of the boundary terms
which arise upon varying the action requires
$\sin (\bh (\Phi - \phi_0 )/2 ) = 0$.   This preserves conformal
invariance, and in terms of the chiral components of the scalar field
implies the `fixed' boundary condition:
\eqn\eIIiii{
{\rm fixed}: ~~~~~
\phi (z) = - \phib (\zb ) + \phi_0 + \frac{2\pi n}{\bh} ~~~~~~~(x=0), }
where
$n$ is an  integer.

\def\psib{\bar{\psi}}

For arbitrary $g$, the theory interpolates between the above free
and fixed boundary conditions.  This model was further studied
in \ref\rskorik{S. Skorik and H. Saleur, {\it Boundary Bound
States and Boundary Bootstrap in the sine-Gordon Model with
Dirichlet Boundary Conditions}, USC-95-01, hep-th/9502011.},
and when $\lambda = 0$ studied
in \ref\rbsw{P. Fendley, H. Saleur and N. P. Warner,
Nucl. Phys. B430 (1994) 577.}.
Interestingly, when $\bh = \sqrt{2} $, the model
is conformally invariant for all $g$ (when $\lambda = 0$)\ref\rcklm{C. Callan,
I. Klebanov, A. W. W. Ludwig and J. M. Maldacena, Nucl. Phys.
B422 (1994) 417.}.

It is well-known that at $\bh = 1$, the bulk theory is equivalent to
free massive charged fermions\ref\rcole{S. Coleman, Phys. Rev. D11 (1975),
2088.}\ref\rmand{S. Mandelstam, Phys. Rev D11 3026 (1975).}.
We now describe how to extend this bosonization to include the
boundary interaction.  Consider first the situation where the bulk
is massless.  Let $\psi_\pm (z) $, $\psib_\pm (\zb )$
be the left and right chiral components of the Dirac fermion with
$U(1)$ charge $\pm 1$.  The bosonisation relations read
$\psi_\pm (z) = \exp (\pm i \phi (z) )$, $\psib_\pm (\zb )
= \exp (\mp i \phib (\zb ) )$.\foot{These bosonisation relations
depend explicitly on the exact form of the Dirac action.
However with these conventions, charge conjugation is
implemented simply by taking $\Phi \rightarrow -\Phi$.}
With these bosonisation relations, $\lambda$ is identified with $-m$, the
soliton mass.
The relative minus sign in
the exponent for left versus right comes from the fact that the
$U(1)$ charge in the massless limit is
 $-i ( \oint \frac{dz}{2\pi i} \d_z \phi  -
 \oint \frac{d\zb}{2\pi i} \d_\zb
 \phib )$.
Thus, in terms of the fermions, the fixed and free boundary conditions
read at $x=0$:
\eqn\eIIiv{\eqalign{
{\rm free}: ~~~~~  \psi_\pm  &= e^{\mp i \sigma} ~ \psib_\mp
{}~~~~~~~  \cr
{\rm fixed}: ~~~~~  \psi_\pm  &= e^{\pm i \phi_0} ~ \psib_\pm
{}~~~~~~~
.\cr}}
One sees that free boundary conditions break $U(1)$ symmetry whereas
fixed boundary conditions preserve it.

We formulate the fermionic action as a perturbation of the
free boundary condition.
The action which enforces the free boundary condition on the fermions
is the following:
\eqn\eIIv{\eqalign{
S_{\rm free} &= -\inv{4\pi} \int dx dt \inv{2} \(\psi_+ \d_\zb \psi_-
+ \psi_- \d_\zb \psi_+
+ \psib_+ \d_z \psib_-  + \psib_- \d_z \psib_+ \) \cr
&~~~~~~~-\frac{i}{8\pi} \int dt \( e^{i\sigma} \psi_+ \psib_+
+ e^{-i\sig} \psi_- \psib_- \)  . \cr}}
Namely, the boundary terms in the variation of the above action
yield \eIIiv.

\def\alb{\bar{\alpha}}

One can now express the boundary perturbation in terms of the
fermion fields.  The chiral components of the free boson have the
following standard mode expansions in the conformal limit:\foot{See
for example \ref\rgins{P. Ginsparg, Les Houches 1988 Lectures,
E. Br\'ezin and J. Zinn-Justin, eds.} }
\eqn\eIIvi{\eqalign{
-i \phi (z) &=   \sum_{n\neq 0} \frac{\al_n}{n} z^{-n}
-\al_0 \log z  - \tilde{\alpha}_0  \cr
-i \phib (\zb) &=   \sum_{n\neq 0} \frac{\bar{\al}_n}{n}
\zb^{-n}
-\alb_0 \log \bar{z}  - \tilde{\alb}_0  \cr
}}
with $[\al_n , \al_m ] = [\alb_n , \alb_m] = n \delta_{n, -m} $,
$[\al_0 , \tilde{\al}_0 ] = [\alb_0 , \tilde{\alb}_0 ] = 1$.
Since we view the interpolating theory as a perturbation of the free
boundary condition, $\phi, \phib$ must satisfy
$\d_z \phi (z) = \d_\zb \phib (\zb ) $ at $x=0$,  which implies
$\al_n = \alb_n$, leaving no constraints on $\tilde{\al}_0 ,
\tilde{\alb}_0$.
On the boundary one has:
\eqn\eIIvii{\eqalign{
\Phi = \phi (z) + \phib (\zb ) &= 2 \phi -i (\tilde{\alb}_0 - \tilde{\al}_0 )
{}~~~~~~~ \cr
&= 2 \phib +i (\tilde{\alb}_0 - \tilde{\al}_0 )
.\cr }}
Therefore at $x=0$,
\eqn\eIIviii{
\cos \( \frac{\Phi - \phi_0}{2} \) =
\( \psi_+ a_-  + \psib_- a_+ \) e^{-i\phi_0 /2}
+ \( a_+ \psi_- + a_- \psib_+ \) e^{i\phi_0 /2 }
}
where $a_\pm$ are zero mode operators:
\eqn\eIIix{
a_\pm = \inv{4} \exp (\pm i (\pi /2 + \phi ( z ) - \bar\phi (\bar z))/2) =
\inv{4} \exp \( \pm (i\pi /2 + \tilde{\al}_0 - \tilde{\alb}_0 )/2 \) . }
The additional factors of $\pi/2$ arise in combining the $\psi$'s and the
$a$'s via $\exp(A) \exp(B) = \exp(1/2[A,B]) \exp(A+B)$.
This leads us to consider the complete action
\eqn\eIIx{
S = S_{\rm free} + S_{\rm mass} + S_{\rm int.}  }
where the interpolating term is
\eqn\eIIxi{\eqalign{
S_{\rm int.}  &= -\frac{g}{4\pi} \int dt  \[
\( \psi_+ a_-  + \psib_- a_+ \) e^{-i\phi_0 /2}
+ \( a_+ \psi_- + a_- \psib_+ \) e^{i\phi_0 /2 } \] \cr
&~~~~~~~~~ - \frac{1}{2\pi}  \int dt ~  \> a_+ \d_t a_-  , \cr}}
and $S_{\rm mass} $ is the bulk mass term
$-\frac{im}{4\pi} \int dx dt (\psi_- \psib_+ - \psib_- \psi_+ )$.
The action \eIIx\ is completely analagous
to the Ising case considered in \rgz.  From the action and the dimension
of the fermion fields one sees that $g$ has dimensions of $\sqrt{\rm mass}$.

Varying the action with respect to the fermion fields and the zero modes
$a_\pm$, the boundary terms, upon elimination of the zero modes, yield
the following interpolating boundary conditions at $x=0$:
\eqn\eIIxii{\eqalign{
\psi_+ + e^{i(\phi_0 - \sigma )} \psi_-
- e^{-i\sigma} \psib_- - e^{i\phi_0 } \psib_+ &= 0 \cr
i\d_t \( \psi_- - e^{i\sigma} \psib_+ \) -  g^2
\( \psib_- e^{-i \phi_0} - \psi_- \) &= 0 \cr
i\d_t \( \psi_+ - e^{-i\sigma} \psib_- \) +  g^2
\( \psi_+  - e^{i\phi_0} \psib_+ \) &= 0  . \cr}}
By taking the $g \rightarrow 0$ ($g\rightarrow \infty$) limit,
we recover the free (fixed) boundary conditions.

\newsec{Derivation of the Boundary S-Matrix}
\def\pap{A^\dagger_a (\theta )}
\def\pbn{A^\dagger_b (-\theta )}
\def\ppp{A^\dagger_+ (\theta )}
\def\ppn{A^\dagger_+ (-\theta )}
\def\pnp{A^\dagger_- (\theta )}
\def\pnn{A^\dagger_- (-\theta )}

\def\Pp{P^+(\theta )}
\def\Pn{P^-(\theta )}
\def\Qp{Q^+(\theta )}
\def\Qn{Q^-(\theta )}

Generally, given a theory with particles $\pap$,
an integrable boundary perturbation
produces scattering via the boundary S-matrix, $R^b_a (\theta )$:
\eqn\eIIIi{\pap B = R^b_a (\theta ) \pbn B ,}
where $B$ is the boundary operator.  In our case, the theory has two particles,
$\ppp$ (solitons) and $\pnp$ (anti-solitons), and the above equation
reads
\eqn\eIIIii{\eqalign{
\ppp B = \Pp\ppn B + \Qp\pnn B ;\cr
\pnp B = \Qn\ppn B + \Pn\pnn B . \cr}}
$\Pp$ and $\Pn$ represent soliton-soliton and anti-soliton-anti-soliton
scattering
whereas $\Qp$ and $\Qn$ represent soliton-anti-soliton scattering.
Because the boundary
interaction, in general, is not U(1) preserving,
the appearance of $\Qp$ and $\Qn$ is expected.
Only when $g \rightarrow \infty$ should $\Qp$ and $\Qn$ vanish.

To derive the pieces of the boundary S-matrix,
we interpret the boundary conditions
in \eIIxii\ to vanish when acting on $B$:
\eqn\eIIIiii{\left[ \psi_+ + e^{i(\phi_0 - \sigma )}\psi_- -
e^{-i\sigma}\psib_- - e^{i\phi_0}\psib_+ \right] B = 0 ~~\hbox{\rm etc.}}
We then substitute  the following mode expansions for the fermions:
\eqn\eIIIiv{\eqalign{
\psi_+ &= \sqrt{m} \int^\infty_{-\infty} {d\theta \over 2\pi i} e^{\theta /2}
\left( A_-(\th ) e^{-m(ze^\th + \bar{z}e^{-\th})}
-\ppp e^{m(ze^\th + \bar{z}e^{-\th})}
\right) ,\cr
\psib_+ &= -i\sqrt{m} \int^\infty_{-\infty} {d\theta \over 2\pi i}
e^{-\theta /2}
\left( A_-(\th ) e^{-m(ze^\th + \bar{z}e^{-\th})}
+\ppp e^{m(ze^\th + \bar{z}e^{-\th})}
\right) ,\cr}}
where $\psi^\dagger_+ = \psi_-$ and $\psib_+^\dagger = \psib_-$.
The $A$'s satisfy the
following anti-commutation relations:
\eqn\eIIIv{\left\{ A_+(\th ),A_+^\dagger (\th ')\right\} = \left\{ A_-(\th ) ,
A_-^\dagger (\th ') \right\} = 4\pi^2\delta (\th - \th ').}
The normalization of the mode expansions
is fixed by insisting the two point functions
(in the bulk) satisfy
\eqn\eIIIvi{\eqalign{
\langle 0|  \psib_- (z,\bar{z}) \psi_+ |0 \rangle = -2im\rm{K}_0 (mr), \cr
\langle 0 | \psi_- (z,\bar{z}) \psi_+  |0 \rangle =
2m\sqrt{\bar{z} \over z}\rm{K}_1 (mr), \cr}}
where $r/2 = \sqrt{z\bar{z}}$ and the K's are standard modified Bessel
functions.  Performing the substitutions, we find
\eqn\eIIIvii{\eqalign{
\Pp &= \left(\cosh (\th ) -
\frac{\gamma}{2} \cosh (\th + i\phi_0) \right) / D(\th ) , \cr
\Pn &= \left(\cosh (\th ) -
\frac{\gamma}{2}  \cosh (\th - i\phi_0) \right) / D(\th ) , \cr
\Qp &= -\frac{i}{2}  e^{-i\sigma} \sinh (2\th ) / D(\th ) , \cr
\Qn &= -\frac{i}{2}  e^{i\sigma} \sinh (2\th ) / D(\th ) , \cr}}
where $D(\th )$ and $\gamma$ are given by
\eqn\eIIIviii{\eqalign{
\gamma &=   g^2/m , \cr
D(\th ) &=
i\gamma\cosh \({\th + i\phi_0 + i\pi /2 \over 2}\)
\sinh \({\th - i\phi_0 -i\pi /2 \over 2} \) -
\cosh^2(\th ) . \cr}}
We see that $\Qp$ and $\Qn$ differ by a phase.
To remove the phase, we apply a U(1) gauge
transformation to $A_+$ and $A_-$:
\eqn\eIIIix{A^\dagger_+ \rightarrow
e^{-i\sigma/2}A^\dagger_+ ~~,~~ A^\dagger_-
\rightarrow e^{i\sigma/2}A^\dagger_- .}
This leaves $\Pp$ and $\Pn$ invariant,
but takes $\Qp \rightarrow e^{i\sigma}\Qp$
and $\Qn \rightarrow e^{-i\sigma}\Qn$.  Then
\eqn\eIIIx{\Qp = \Qn = -\frac{i}{2} \sinh (2\th ) / D(\th ) . }
In what follows, we assume this gauge choice has been made.

If $\phi_0 = 0$, we expect charge conjugation symmetry to be restored,
as indeed it is.
We also see that the R-matrix is independent of the sign of g.
Changing the sign of g is equivalent to the shift $\phi_0 \rightarrow
\phi_0 + 2\pi$.  But the subsequent
shift $\Phi \rightarrow \Phi + 2\pi$ restores the boundary term and leaves
the bulk terms unchanged, leading to the independence from the sign of
g.

Taking the appropriate limits,
we can find the boundary S-matrix in the free and fixed
cases.  For the free case we find
\eqn\eIIIxi{ {\rm free}: ~~~~~
P^\pm (\th ) = - \rm{sech} (\th ), ~~~~~~~~~~~~~~~~
Q^\pm (\th ) = i\tanh (\th). }
In this limit,
charge conjugation symmetry, $\Phi \to - \Phi$, is respected.  As
a result, $\Pp$ equals $\Pn$.  For the fixed case we obtain
\eqn\eIIIxii{ {\rm fixed}:~~~~~
P^\pm (\th ) = -\cosh \({\th \pm i\phi_0 -i\pi/2 \over 2}\) /
\cosh\({\th \mp i\phi_0 +i\pi/2 \over 2} \) , ~~~~~~
Q^\pm (\th ) = 0 . }
As expected the soliton-antisoliton
scattering amplitudes vanish in this limit where the
boundary interaction preserves the U(1) charge.

\newsec{Analytic Structure of the Boundary S-Matrix}

Like the bulk S-matrix,
the poles of the boundary S-matrix, R, provide information on bound
states.  However in the case of R,
the situation is more complicated.  Poles in R can
either be indicative of bulk bound states
interacting with the boundary or of boundary
bound states, which are effectively excitations of the ground boundary
state.  In the former
case, poles in R will both be found at
$\th = i\pi/2$ and at $i\pi/2 - \th_b$, where
$\th_b$ is the location of the pole in S
corresponding to the bulk bound state.  Such poles
necessarily imply the addition
of zero-momentum states to the boundary state  $|B\rangle$.
In the latter case,
the poles may appear anywhere in the region
$0 \leq \th \leq i\pi/2$.  If the pole in this case appears at
$\th = i\pi/2$, a zero-momentum state will, as before, be found in
$|B\rangle$.
Because the bulk S-matrix at
the free-fermion point
has no pole structure, we are only faced with the latter situation.

\def\gam{\frac{\gamma}{2}}

The poles in the range
$0 < \th < i\pi/2$ appearing in the free-fermion boundary S-matrix
depend on the parameters $\gamma$ and $\phi_0$.  There are four areas in this
parameter space:
\eqn\eIVi{\eqalign{
{}~~~~~~~~~i)&~ \frac{\gamma}{2} \cos\phi_0 > 1,~ \cos{\phi_0}<1: ~~~
u = \sin^{-1}\left(\gamma/4 - (\gamma^2/16 - \gam\cos\phi_0 + 1)^{1/2}
\right) ;\cr
ii)&~\gam\cos\phi_0 > 1,~ \cos{\phi_0}=1,~\gamma<4:
{}~~~u=\sin^{-1}(\gamma/2-1);\cr
iii)&~\gam\cos\phi_0 > 1,~ \cos{\phi_0}=1,~\gamma \geq 4: ~~~
\rm{no~poles};\cr
iv)&~ 0 < \gam\cos\phi_0 \leq 1: ~~~\rm{no~poles} ; \cr}}
where $u = -i\th$
is the location of the poles.  In the free and fixed cases, this pole
structure reduces to
\eqn\eIVii{\eqalign{
{}~~~~~&\rm{fixed:} ~~~ u = \pi/2 - |\phi_0 |  ~,~ ~~~0< |\phi_0|  <\pi/2 ;\cr
{}~~~~~&\rm{free:} ~~ u = \pi/2 . \cr}}
In all cases the poles are simple, and
we are not burdened with interpreting more complicated
resonance phenomena.

The energy, $E_\alpha$, of  these boundary
bound states, $B_\alpha$, is given by
\eqn\eIViii{E_\alpha - E_0 = m\cos (u_\alpha ),}
where $m$ is the soliton mass, $E_0$ is
the energy of the boundary ground state, and
$u_\alpha$ is the location of the pole in
the boundary S-matrix corresponding to the state.
We expect the boundary bound state $B_\alpha$ to
be stable if $E_\alpha - E_0 < m$; that
is, if there is no chance of the boundary
emitting a zero-momentum soliton.  We thus see
all the boundary states to be stable.

When boundary bound states are present,
scattering off the boundary \eIIIi\ needs
to be described by the more general equation
\eqn\eIViii{\pap B_\alpha = R^{b\beta}_{a\alpha} (\th ) \pbn B_\beta ,}
where $\{B_\alpha\}$ are all the possible boundary states.
By energy conservation, we
expect $R^{a\alpha}_{b\beta} = 0$
if $E_\alpha \neq E_\beta$.  Thus for all but the free
case and the fixed case with $\phi_0 = 0$, the
boundary states are not coupled.
With such independence, we expect $R^{b\alpha}_{a\alpha}
= R^{b\beta}_{a\beta}$
as the constraints among the free-fermi fields \eIIxii\ must
annihilate $B_\alpha$ for every $\alpha$.

The interpretation of the boundary states in these non-degenerate
cases is straightforward.
Classically, the asymptotic behaviour of the ground state in the bulk
is described by
$\Phi \rightarrow (2n+1)\pi$ as $x \rightarrow -\infty$ (the bulk classical
potential is $V(\Phi ) = \frac{|\lambda |}{8\pi} \cos (\Phi )$; recall
$\lambda = -m$ and so $\lambda$ is negative).
The difference in energy between these
states arises from a boundary term.  Provided $\gamma$ is large enough,
the boundary term forces $\Phi \sim \phi_0$ at $x=0$.
If $\phi_0$ is in the range $0 < \phi_0 < \pi/2$, the boundary bound state can
be viewed as the state where $\Phi$'s
asymptotic value is $-\pi$ (for the ground state the asymptotic value of
$\Phi$ is $\pi$).  If $-\pi /2 < \phi_0 < 0$ the situation is reversed:
the ground state corresponds to $\Phi \rightarrow -\pi$ and the excited
boundary state to $\Phi \rightarrow \pi$ as $x \rightarrow -\infty$.
If however $\pi/2 < \phi_0 < \pi$, there is no boundary excited state:
$\Phi$ is too near either of the bulk vacuua values $\pm\pi$ at the boundary
to drift over to the other as $x \rightarrow -\infty$.

In the fixed degenerate case (i.e. $\phi_0 = 0$), the interpretation is
much the same.
$\phi_0 = 0$ is exactly between the bulk ground state
field configurations $\Phi = \pm \pi$.  Thus $\Phi \rightarrow
\pm \pi$ as $x \rightarrow -\infty$ are energetically equivalent.
Because of this degeneracy there is the possibility that
$R^{b\beta}_{a\alpha}$ does not vanish if $\alpha \neq \beta$.  To derive
these cross-scattering matrix elements, one would presumably solve
\eqn\eiViii{f(\psi_{\pm},\bar\psi_{\pm} ) B_\alpha =
\sum_{\beta\neq\alpha} R^{b\beta}_{a\alpha} (\th ) A^\dagger_b (-\th )
B_\beta ,}
where $f$ is a combination of fields appearing in \eIIxii\ . Thus the
matrix elements diagonal in $\alpha$ and $\beta$ remain unchanged.  With
the presence of scatttering between boundary states, the unitarity condition
that must be satisfied is
\eqn\eiViv{R^{b\beta}_{a\alpha} (\th ) R^{c\gamma }_{b \beta } (-\th )
= \delta_{ac}\delta_{\alpha\gamma} .}
Given that we know in our case (as is easily checked)
the diagonal matrix elements satisfy unitarity by
themselves, it is straightforward to check that unitarity forces
$R^{a\alpha}_{b\beta} \propto \delta_{\alpha\beta}$.  Thus we still have
no scattering between boundary states.  The construction of the two boundary
states $B_\alpha$ in this case
will include a zero momentum soliton (if the pole
appears in $P_+$) or anti-soliton (if the pole appears in $P_-$).

In the free (degenerate) case the interpretation of the boundary state differs.
Considering the massless limit,
the twofold degenerate ground states are characterized by asymptotic
behaviour $\Phi \rightarrow \pm\pi$ as
$t \rightarrow \infty$
and $\Phi \rightarrow \mp \pi$ as $t \rightarrow -\infty$.
These two regions of different regions of $\Phi$ are separated by a domain
wall which can be thought of as a zero momentum particle present in the
boundary state.  This description is analogous to the description of
semi-infinite Ising model at criticality in \rgz .

Our discussion has not yet exhausted
all the poles in the physical strip $0 < \th < i\pi$.
For every pole in the range $0 < \th < i\pi/2$, there is a corresponding
pole in the
range $i\pi/2 < \th < i\pi$.
These additional poles correspond to scattering in the cross
channel.

\newsec{Matching Ghoshal and Zamolodchikov's Solution}
We can now match our
boundary S-matrix to that of Ghoshal and Zamolodchikov \rgz, thereby
relating the formal parameters in the
latter to the physical parameters in the action \eIIx\
.  At the free-fermion point
(corresponding to the limit $\lambda \rightarrow 1$ in the
notation of \rgz ), the solution of the
boundary Yang-Baxter equation yields
\eqn\eVi{\eqalign{
\Pp &= \cos (\xi - i\th ) R(\th ) ;\cr
\Pn &= \cos (\xi + i\th ) R(\th ) ;\cr
\Qp &= -i\frac{k_+}{2} \sinh (2\th ) R(\th );\cr
\Qn &= -i\frac{k_-}{2} \sinh (2\th ) R(\th );\cr}}
where $k_\pm$ and $\xi$ are free parameters.
Using the aforementioned gauge transformation,
we set $k_+ = k_- = k$.  From the results of the last section
we thus see that the phase $\sigma$ removed from $k_\pm$ in \rgz\
is nothing other than the phase which specifies the conformal
free boundary condition.

The function $R(\th )$ is
determined through the boundary unitarity and boundary
cross-unitarity equations.  In the free-fermion limit these constraints yield
\eqn\eVii{R(\th ) = {1\over \cos \xi }
\sigma(\eta , -i\theta )\sigma (i\vartheta , -i\theta ), }
where the parameters $\eta$ and $\vartheta$
are related to $k$ and $\xi$ through the equations
\eqn\eViii{\eqalign{
\cos (\eta )\cosh (\vartheta ) &= -{1\over k} \cos(\xi ) ;\cr
\cos^2(\eta ) + \cosh^2(\vartheta ) &= 1 + {1\over k^2} . \cr}}
The function $\sigma (x,u)$ is found to be
\foot{The denominator of (5.23) in \rgz ~
should be $\Pi^2(x, \pi/2)\Pi^2(-x , -\pi/2)$,
and (5.24) should read
$\sigma (x, u) \sigma (x, -u) = \cos^2 x  [\cos (x + \lambda u )
\cos (x-\lambda u ) ]^{-1} $.}
\eqn\eViv{\sigma (x,u) = \cos x
\left(2 \cos (\pi/4 + x/2 - u/2) \cos (\pi/4 - x/2 - u/2)
\right)^{-1}.}

With some algebra one can show
that this solution of Ghoshal and Zamolodchikov maps onto
our derivation of the boundary scattering matrix with the following
identification of the free parameters:
\eqn\eVv{\eqalign{
k &= \left(1 - \gamma\cos (\phi_0 ) + \gamma^2/4\right)^{-1/2} ; \cr
\cos \xi &= k\left( 1-\frac{\gamma}{2} \cos \phi_0 \right) ; \cr
\sin \xi &= -k \frac{\gamma}{2} \sin \phi_0  . \cr}}
In the limit $\gamma \rightarrow 0$,
we see that $k=1$ and $\xi = 0$ yielding in Ghoshal
and Zamolodchikov's solution, $\Pp = \Pn$ and $\Qp = \Qn$,
in accordance with charge
conjugation symmetry. In the limit of fixed boundary condition,
$\gamma \rightarrow \infty$
we see that $k=0$  and
\eqn\eVvi{\xi = \phi_0 + (2n+1)\pi ,}
forcing their amplitudes $Q_{\pm}$
to vanish in accordance with U(1) charge conservation.  We point out that
the relationship between $\xi$ and $\phi_0$ is not in exact accordance with
that conjectured in
\rgz ~(see 5.27).
There, it was taken  that when $\phi_0$ vanished, $\xi$ was forced to vanish
in order to ensure $\Pp = \Pn$.
But $\xi = \pi$ also sets $\Pp = \Pn$.

It should be emphasized that \eVvi\
depends strongly on the conventions used in writing the bulk Dirac action.
Taking $\gamma_\mu \rightarrow -\gamma_\mu$ (where $\gamma_\mu$ are the
gamma matrices used in the action) leads to the relationship as conjectured
in \rgz , namely
\eqn\eVvii{\xi = \phi_0 + 2n\pi .}
In this case $\lambda = m$ and the poles in $R(\th )$ shift to reflect the
ground states in the bulk are now $\Phi = 2n\pi$.

For both \eVvi\ and \eVvii\ , taking $\Phi \rightarrow -\Phi$ implements charge
conjugation.  However it is easy enough to construct Dirac actions where this
action does not implement charge conjugation exactly, for example where the
bosonisation relations are $\psi_\pm = e^{\pm i\phi}$, $\bar\psi_\pm =
\pm i e^{\mp i \bar\phi}$.  In this case the relation between $\phi_0$ and
$\xi$ takes the form $\phi_0 = \xi \pm \pi/2$.  Conventions of this sort
are used in
\rskorik.  However these authors
have taken the relation between $\xi$ and
$\phi_0$ as given in \eVvii\ .  This has led them to find a `boundary
anomaly'.  This anomaly should vanish when the correct relation corresponding
to their conventions between
$\phi_0$ and $\xi$ is used.

\bigskip

\centerline{Acknowledgements}

This work is supported by an Alfred P. Sloan Foundation fellowship,
and the National Science Foundation in part through the National
Young Investigator program.


\listrefs

\bye